\documentclass[showkeys]{revtex4}
\usepackage{graphicx}
\usepackage{dcolumn}
\usepackage{amsmath}
\usepackage{color}
\usepackage{epstopdf}
\usepackage{float}
\usepackage{graphicx}
\usepackage{subfigure}
\usepackage[utf8x]{inputenc}
\definecolor{coolblack}{rgb}{0.0, 0.18, 0.39}
\definecolor{darkred}{rgb}{0.5,0,0}
\definecolor{darkgreen}{rgb}{0,0.5,0}
\definecolor{darkblue}{rgb}{0,0,0.5}
\definecolor{lapislazuli}{rgb}{0.15, 0.38, 0.61}
\definecolor{venetianred}{rgb}{0.78, 0.03, 0.08}
\definecolor{bleudefrance}{rgb}{0.19, 0.55, 0.91}
\definecolor{dogwoodrose}{rgb}{0.84, 0.09, 0.41}
\usepackage[linktocpage,colorlinks]{hyperref}
\hypersetup{colorlinks=true, citecolor=darkgreen, linkcolor=blue,urlcolor = darkblue}
\makeatletter
\def\btt#1{\texttt{\@backslashchar#1}}
\DeclareRobustCommand\bblash{\btt{\@backslashchar}} \makeatother

\RequirePackage{fix-cm}

\begin{document}
\title{Stability of circular geodesics in equatorial plane of Kerr spacetime %\thanksref{t1}  
}
\author{Pradeep Singh $^{a}$}\email{pradeep.dawan@gmail.com}
\author{ Hemwati Nandan $^{b,c}$}\email{hnandan@associates.iucaa.in}
\author{Lokesh Kumar Joshi $^{d}$}\email{lokesh.joshe@gmail.com}
\author{Nidhi Handa $^{a}$}\email{handanidhi1@gmail.com}
\author{Shobhit Giri $^{b}$}\email{shobhit6794@gmail.com}
\affiliation{$^{a}$Department of Mathematics \& Statistics, Gurukula Kangri (Deemed to be University), Haridwar 249 404, Uttarakhand, India.}
\affiliation{$^{b}$Department of Physics, Gurukula Kangri (Deemed to be University), Haridwar 249 404, Uttarakhand, India}
\affiliation{$^{c}$Center for Space Research, North-West University, Mahikeng 2745, South Africa}
\affiliation{$^{d}$ Department of Applied Science, Faculty of Engineering and Technology, Gurukula Kangri (Deemed to be University), Haridwar 249 404, Uttarakhand, India}

% The correct dates will be entered by the editor

\begin{abstract}
\noindent We analyze the  stability of circular geodesics for timelike as well as null geodesics of the Kerr BH spacetime with rotation parameter on the equatorial plane by Lyapunov stability analysis. Also, we verify the results of stability by presenting the phase portrait for both timelike and null geodesics. Further, by reviewing the Kosambi-Cartan-Chern (KCC) theory, we analyze the Jacobi stability for Kerr spacetime and present a comparative study of the methods used for stability analysis of geodesics. 
% \PACS{PACS code1 \and PACS code2 \and more}
% \subclass{MSC code1 \and MSC code2 \and more}
\end{abstract}
\maketitle
\section{Introduction}
%%%%%%%%%%%%%%%%%%%%%%%%%%
\noindent The generalized black hole (BH) solution in general relativity (GR) that includes spin along with mass is known as Kerr BH. This BH was observed as an exact solution of Einstein field equations whose mathematical description  is given by Roy Kerr \cite{hartle2003gravity,Wald1984c}. This BH solution is different from Schwarzschild BH (SBH) solution in many aspects, such as it is spinning BH having two symmetries; one in time and the other along a certain axis. The Kerr metric is parameterized by its angular momentum and mass while SBH metric is only parameterized by their mass and it has only spherical symmetry \cite{chandrasekhar1998mathematical,    abolghasem2013stability,weber2018kerr}. The motion of test particles in curved spacetime under the influence of gravity is governed by geodesic equations \cite{hartle2003gravity,fujita2009analytical,marck1996short}. As the mathematical formulation of geodesic equations in Kerr spacetime are non-linear differential equation which are difficult to solve explicitly. To overcome this difficulty, we use the Lyapunov stability analysis of non linear system along with phase portraits \cite{hackmann2010geodesic,lakshmikantham1989stability,giri2021stability,cardoso2009geodesic}. A. M. Lyapunov first developed stability theory for non-linear ordinary differential equations by characterizing the behavior of the dynamical systems trajectories on the basis of nearby solutions \cite{sastry1999lyapunov,goldhirsch1987stability}.\\ The main aim of our analysis of dynamical system is that, whether or not the system has stable equilibria. We characterize an equilibrium as stable or unstable based on the  behavior of solutions whose initial conditions are in the neighborhood of the equilibrium \cite{eberly2008stability, roussel2005stability}. As phase portraits combined with linear stability analysis can generally provide us a full picture of the dynamics. However, it becomes much more difficult in higher-dimensional spaces so one can use the Lyapunov stability analysis technique to determine the stability of an equilibrium point  both near and far from the equilibrium point. The stability analysis carried out at each equilibrium point involves Lyapunov's linear stability analysis and Jacobi stability analysis method \cite{sastry1999lyapunov,boehmer2012jacobi,sandri1996numerical}. Jacobi stability analysis is carried out by using Kosambi-Cartan-Chern (KCC) theory which is widely used to investigate the properties of dynamical system in terms of five geometrical invariants. The second KCC invariants which is also known as a deviation curvature tensor gives us the Jacobi stability of the trajectories which measures the robustness of the second-order differential equation \cite{harko2016kosambi}. The Jacobi stability studies the robustness of second order differential equation which is analyzed by calculating deviation curvature tensor (second KCC invariant) by  KCC theory \cite{yajima2008nonlinear,gupta2017kcc,abolghasem2013jacobi,abolghasem2012liapunov}. The Lyapunov and Jacobi stability of circular orbits in the SBH spacetime has already been investigated in detail by Hossein \cite{abolghasem2013stability}. However, various investigations regarding geodesics stability around BH spacetimes have been performed time and again in diverse context \cite{cardoso2009geodesic,acena2020circular,giri2021stability}. Here our main objective is to perform the stability analysis of geodesics of Kerr BH with a different approach namely the Lyapunov and Jacobi stability analysis to have a more precise information regarding the stability of circular geodesics which are important from the view point of astrophysics. \\
The present paper is organized as follows. In section 2, we review the basic mathematical concepts of the stability theory of the dynamical system. In section 3, we introduce the metric of rotating BH and Euler-Lagrange equation are used to obtain geodesic equations. We have also studied the variation of effective potential with rotation parameter. In section 4, the  Lyapunov stability for timelike as well as null geodesics is calculated at the equilibrium point. Also, phase portrait in $r-p$ plane for timelike as well as null geodesics are presented in this section. In section 5, we analyze the Jacobi stability of the system by using KCC theory. Finally, in section 6, we discuss and conclude the obtained results.     
\section{Stability analysis of dynamical system}
In this section, we first review the equilibrium point of the differential equation and its classification with its nature.
We also review some mathematical concepts of stability of dynamical system used in this work \cite{sedin2016stability}. As the dynamical system may be linear or non-linear, some non-linear systems can be very complicated and difficult to solve explicitly. To analyze such systems, we first linearize them at their equilibria and then construct a phase portrait to visualize the trajectories of the solutions of the system  \cite{robinson1998dynamical,morgan2015linearization}.
\subsection{Classifying equilibria with stability}
In a differential equation, an equilibrium point is a constant solution of the equation. For a system of autonomous ODE's 
\begin{equation}
\dot{X}=f\left(X\right),
\end{equation}
a point $X^{*}\in\ R^{n}$ will be an equilibrium point if $f(X^{*})=0$. This equilibrium point is called hyperbolic if all the eigen values of the Jacobian matrix evaluated at equilibrium point have a negative  real part otherwise the point is non-hyperbolic.
If $\lambda_{1}$ and $\lambda_{2}$ are the eigen values of the Jacobian matrix $J$ evaluated at the equilibrium point $X^{*}$ of the two dimensional system of differential equation then the nature and stability of equilibrium points are classified as shown in \tablename{\ref{stabilitytable}}.
\begin{table*}[ht]
	\caption{Classification of equilibrium point with their stability by using the Lyapunov's linear stability analysis \cite{boehmer2012jacobi}}
	\centering
	\begin{tabular}{|c|c|c|}
		\hline
		Eigen values of Jacobian matrix &  Types of equilibrium point & Stability   \\  [0.5ex]
		\hline
		$\lambda_{1}\leq \lambda_{2}<0$ & Node$\left(sink\right)$ & Stable \\
		
		$\lambda_{1}\geq \lambda_{2}>0$  & Node$\left(source\right)$ & Unstable \\
		
		$\lambda_{1}<0< \lambda_{2}$ & Saddle & Unstable \\
		$\lambda_{1,2}=a\pm ib$ with $a<0$ & Spiral & Stable \\
		$\lambda_{1,2}=a\pm ib$ with $a>0$ & Spiral & Unstable \\
		$\lambda_{1,2}=a\pm ib$ with $a=0$ & Center & Stable \\
		\hline
	\end{tabular}  
	\label{stabilitytable}
\end{table*}
\\    Boehmar et al.\cite{boehmer2012jacobi} broadly represented the phase portraits of a two dimensional differential equation for the various cases of the eigenvalues of the Jacobian matrix of the system. 
\subsection{Jacobi stability via KCC theory}
To investigate the properties of the dynamical systems, a powerful mathematical theory was proposed by the Kosambi-Cartan-Chern (KCC) \cite{kosambi2016parallelism,cartan2016observations,chern1952elie}. In this theory, the properties of any dynamical system are described in terms of five geometrical invariants with the second one giving the Jacobi stability of the system.
Consider a system of second order differential equation \cite{harko2015new}
\begin{equation}
\frac{d^{2}x^{i}}{dt^{2}}+2G^{i}{\left(X,Y\right)}=0 ; i=1,2,3,...,n\label{general kcc eq},
\end{equation} 
where $G^{i}\left(X,Y\right)$ are smooth functions defined in a local system of coordinates on TM(tangent bundle of real, smooth n-dimensional manifold M) with $X=\left(x^{1},x^{2},...x^{n}\right), Y=\left(y^{1},y^{2},...y^{n}\right)$.\\	The second KCC invariant which is also known as a deviation curvature tensor of the system of second order differential equation\eqref{general kcc eq} is given by
\begin{equation}
P^{i}_{j}=-2\frac{\partial G^{i}}{\partial x^{j}}-2G^{l}G^{i}_{jl}+y^{l}\frac{\partial N^{i}_{j}}{\partial x^{l}}+N^{i}_{l}N^{l}_{j},
\end{equation}
where $G^{i}_{jl}=\frac{\partial N^{i}_{j}}{\partial y^{l}}$ is called the Berward connection and $N^{i}_{j}=\frac{\partial G^{i}}{\partial y^{j}}$ defines the coefficient of a non linear connection N on the tangent bundle TM.\\
The trajectories of given system of second order differential equations are Jacobi stable if and only if the real parts of the eigen values of the deviation curvature tensor $P^{i}_{j}$ are strictly negative everywhere, and Jacobi unstable otherwise. Geometrically, the trajectories of the system are bunching together if they are Jacobi stable and dispersing if they are Jacobi unstable.
\section{The Kerr BH spacetime}
The Kerr BH spacetime is an exact solution of Einstein's field equations in vacuum describing an axisymmetric, non-static, asymptotically flat gravitational field \cite{visser2007kerr,pugliese2011equatorial}.  In Boyer-Lindquist coordinates the geometry of Kerr metric is given as follows  
\begin{multline}
ds^{2}=g_{ab}dx^{a}dx^{b}=-\left(1-\frac{2Mr}{\rho^{2}}\right) dt^{2} -\left(\frac{4Mra\sin^{2}\theta}{\rho^{2}}\right) d\phi dt+\frac{\rho^{2}}{\Delta} dr^{2}+\rho^{2} d\theta^{2} \\+\left(r^{2}+a^{2}+\frac{2Mra^{2}\sin^{2}\theta}{\rho^{2}}\right)\sin^{2}\theta d\phi^{2}.
\end{multline}
where~~ $\rho^{2}=r^{2}+a^{2}\cos^{2}\theta$,~~ and ~$\Delta=r^{2}-2Mr+a^{2}$.\\Here the specific angular momentum (Kerr parameter) is defined as $a=J/M$, where J is angular momentum and M is the mass of the BH. The coordinates $\left(t,r,\theta,\phi\right)$ used in this geometry are called Boyer-Lindquist coordinate. To analyze the stability of geodesics in Kerr BH spacetime on  the equatorial plane, we fixed  $\theta=\pi/2$, for which $\rho=r$ and~~ $d\theta=0$. 
Thus, the geometry of Kerr metric is reduces to,
\begin{equation}
ds^{2}=-\left(1-\frac{2M}{r}\right)dt^{2}-\frac{4aM}{r} dt d\phi+\frac{r^{2}}{\Delta} dr^{2}+  
\left(r^{2}+a^{2}+\frac{2Ma^{2}}{r}\right)d\phi^{2},
\end{equation}
From the Euler-Lagrange equation the generalized momentum for the geodesics of Kerr BH are as follows
\begin{equation}
\dot{t}=\frac{1}{\Delta}\left[\left(r^2+a^2+\frac{2Ma^2}{r}\right)e-\frac{2Mal}{r}\right],
\end{equation}
\begin{equation}
\dot{\phi}=\frac{1}{\Delta}\left[\left(1-\frac{2M}{r}\right)l+\frac{2Mae}{r}\right],
\end{equation}
\begin{equation}
\dot{r}=\sqrt{e^{2}+K-\frac{2KM}{r}-\frac{l^{2}-a^2\left(e^2+K\right)}{r^{2}}+\frac{2 M\left(l-ae\right)^{2}}{r^{3}}} \label{radialeq},
\end{equation}
where,
\begin{equation}
K =
\begin{cases}
\ -1, \text{ for  timelike  geodesics} \\
\  0, \text{ for null geodesics}
\end{cases}
\end{equation}
Here $e$ and $l$ represent the energy and angular momentum per unit mass of the particle moving around BH respectively.\\ Now from the Eq. \eqref{radialeq}  we obtain, radial equation as
\begin{equation}
\dot{r}^2=e^{2}+K-\frac{2KM}{r}-\frac{l^{2}-a^2\left(e^2+K\right)}{r^{2}}+\frac{2 M\left(l-ae\right)^{2}}{r^{3}}.
\end{equation}
This equation can be written in the following form
\begin{equation}
\frac{e^{2}}{2}=\frac{\dot{r}^{2}}{2}+V\left(r\right),\label{energyeq} 
\end{equation}
where, 
\begin{equation}
V\left(r\right)=-\frac{K}{2}+\frac{KM}{r}+\frac{l^{2}-a^{2}\left(e^{2}+K\right)}{2r^{2}}-\frac{M\left(l-ae\right)^{2}}{r^{3}} \label{Kerr effective potential eq}
\end{equation}
is the expression of effective potential for   Kerr BH spacetime. By putting the values of $K$ in Eq. \eqref{Kerr effective potential eq}, we can obtained the effective potential expression for timelike and null geodesics. The graphical representation of effective potential is shown in the figure given below for the various values of $'a'$.
\begin{figure}[H] 
\centering
	\subfigure[]{\includegraphics[width=8cm,height=8cm]{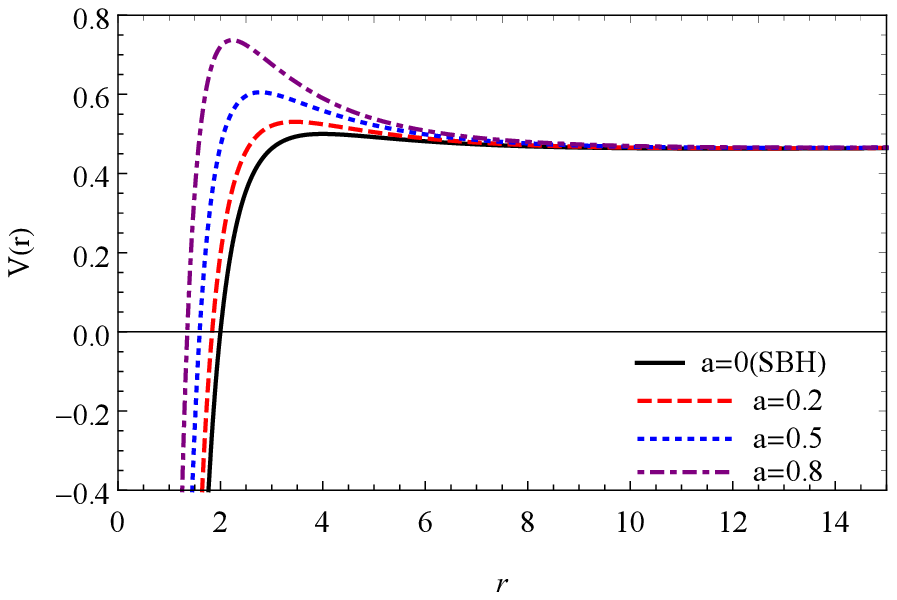}\label{effpottime}}
	\subfigure[]{\includegraphics[width=8cm,height=8cm]{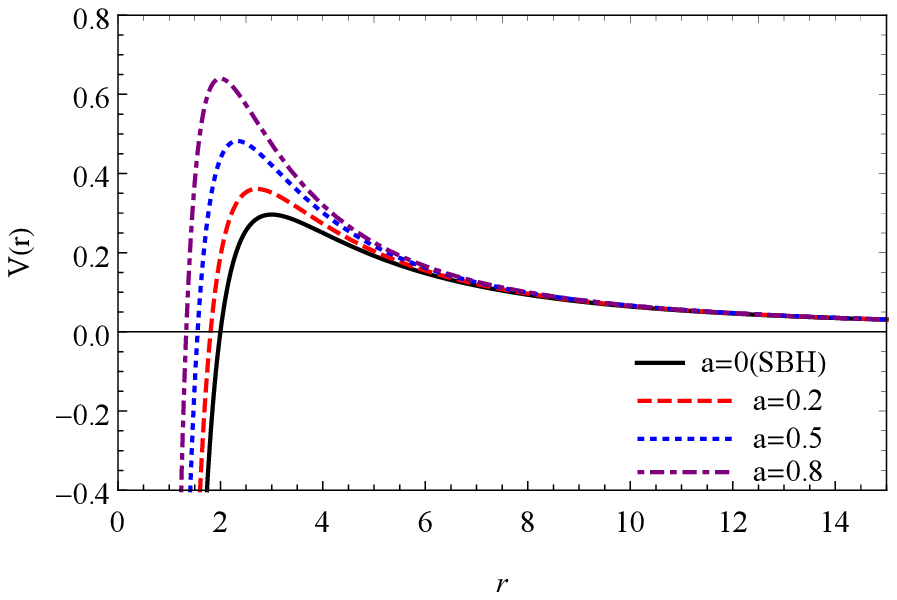}\label{effpotnull}}
	\caption{The effective potential as a function of $r$ with various values of Kerr parameter $'a'$ for (a) timelike and (b) null geodesics.}
\end{figure}
To study the behavior of the test particle near BH, \figurename{\ref{effpottime}} and \figurename{\ref{effpotnull}} shows graphical representation of effective potential for timelike and null geodesics for various value of Kerr parameter $'a'$. The same is also presented for SBH case with limit $a=0$.  In each curve of  Fig.1(a), we see that there are two extreme points in which one is maxima and other is minima. Thus, stable circular orbits exist correspond to point of minima, while in each curve of Fig.1(b) there is no minima which indicates no stable circular orbits exists in case of null geodesics. 
\section{Lyapunov stability analysis}
The derivative of \eqref{energyeq} w.r.t. affine parameter $\tau$ gives us a one-dimensional geodesics equation
\begin{equation}
\ddot{r}=-V'\left(r\right). \label{one-dim.geodesiceq}
\end{equation}
Then Eq. \eqref{one-dim.geodesiceq} can be transformed into the system of first order differential equation in $r-p$ space as follows,
\begin{equation}
\begin{cases}
\  \dot{r}=p, \\
\dot{p}=-V'\left(r\right).
\end{cases} \label{one-dim system}
\end{equation}
To apply linearization process, let $f:\left(0,\infty\right)\times R\rightarrow R^2$ be the vector field $f\left(r,p\right)=\left(p,-V'\left(r\right)\right)$ for the system \eqref{one-dim system}.\\
Clearly, the equilibrium point of this system are the points $\left(r_{*}, 0\right)$, where $r_{*}$ is the solution of the equation $V'\left(r\right)=0$. Now the Jacobian matrix for the above system at the any point $\left(r, p\right)$ is defined as,
\begin{equation}\label{key}
J= \frac{\partial f}{\partial x}\left(r,p\right)=
\begin{bmatrix}
0 & 1 \\
-V''(r) & 0
\end{bmatrix}.
\end{equation}
Now the eigenvalues obtained from the characteristic equation of this Jacobian matrix at the equilibrium point $\left(r_{*}, 0\right)$  are given as,
\begin{equation}
\lambda=\pm\sqrt{-V''\left(r_{*}\right)} \label{ch.eq}.
\end{equation}
Thus the equilibrium point is a saddle point when $V''\left(r_{*}\right)<0$ and is a possible center when $V''\left(r_{*}\right)>0$. The Lyapunov function for the system is defined as
\begin{equation}
E=\frac{p^{2}}{2}+V\left(r\right),
\end{equation} 
and its Hessian matrix is as follows
\begin{equation}\label{key}
H=\frac{\partial^{2}E}{\partial r\partial p}=
\begin{bmatrix}
V''(r_{*}) & 0 \\
0 & 1
\end{bmatrix}.
\end{equation}
Clearly this matrix is positive definite when $V''\left(r\right)>0$, and thus at a possible center $\left(r_{*}, 0\right)$ the Lyapunov function has a local minimum. Thus the equilibrium point  $\left(r_{*}, 0\right)$ is said to be
\begin{equation}
\begin{cases}
\ \text{Lyapunov stable if},  V''\left(r_{*}\right)>0,\\
\ \text{Lyapunov unstable if}, V''\left(r_{*}\right)<0.
\end{cases}
\end{equation}
\subsection{For timelike geodesics}
\begin{figure} [H]
	\subfigure[]{\includegraphics[width=09cm,height=7.5cm]{pptime1.eps}}
	\subfigure[]{\includegraphics[width=09cm,height=7.5cm]{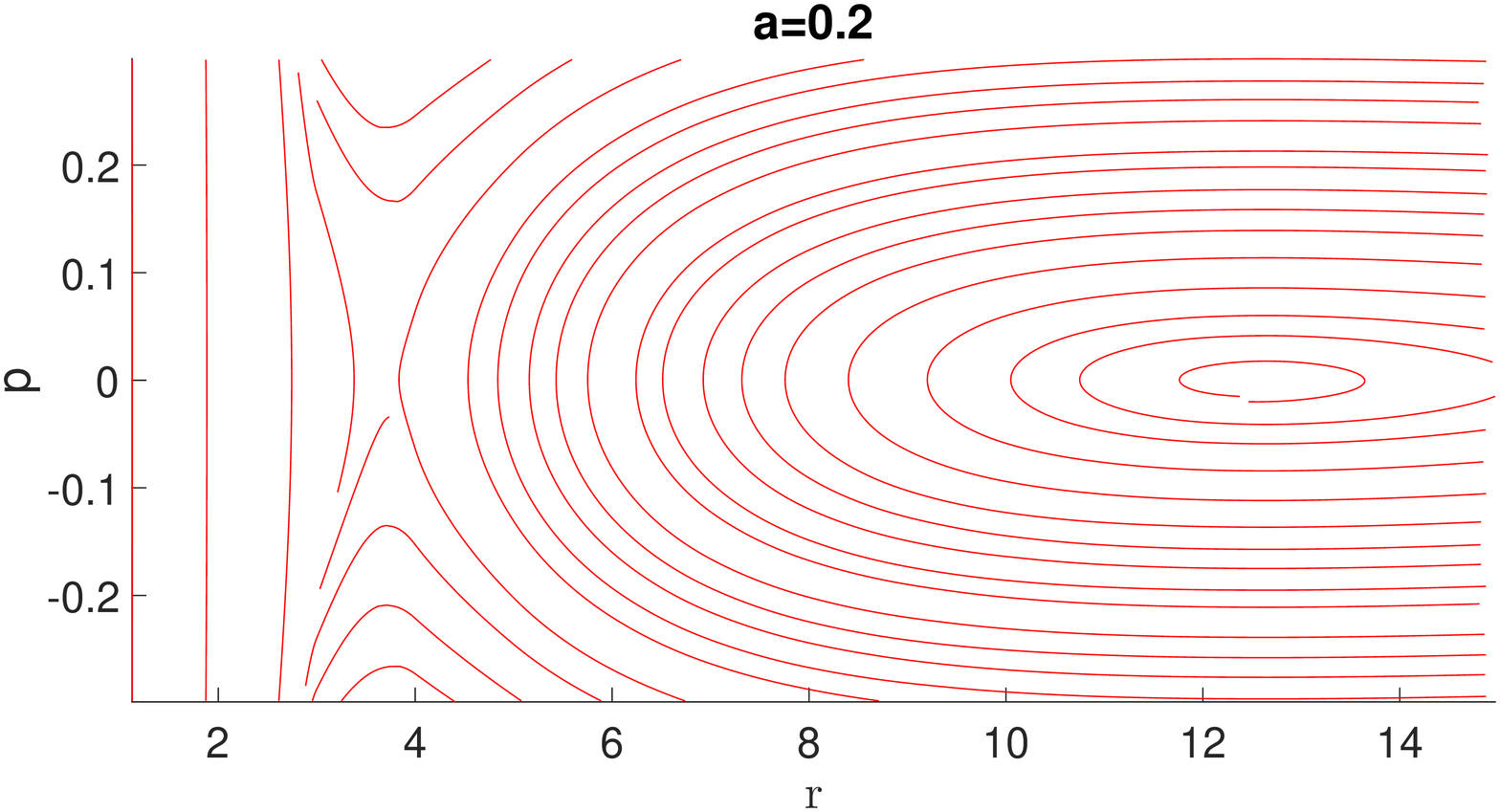}}
	\subfigure[]{\includegraphics[width=09cm,height=7.5cm]{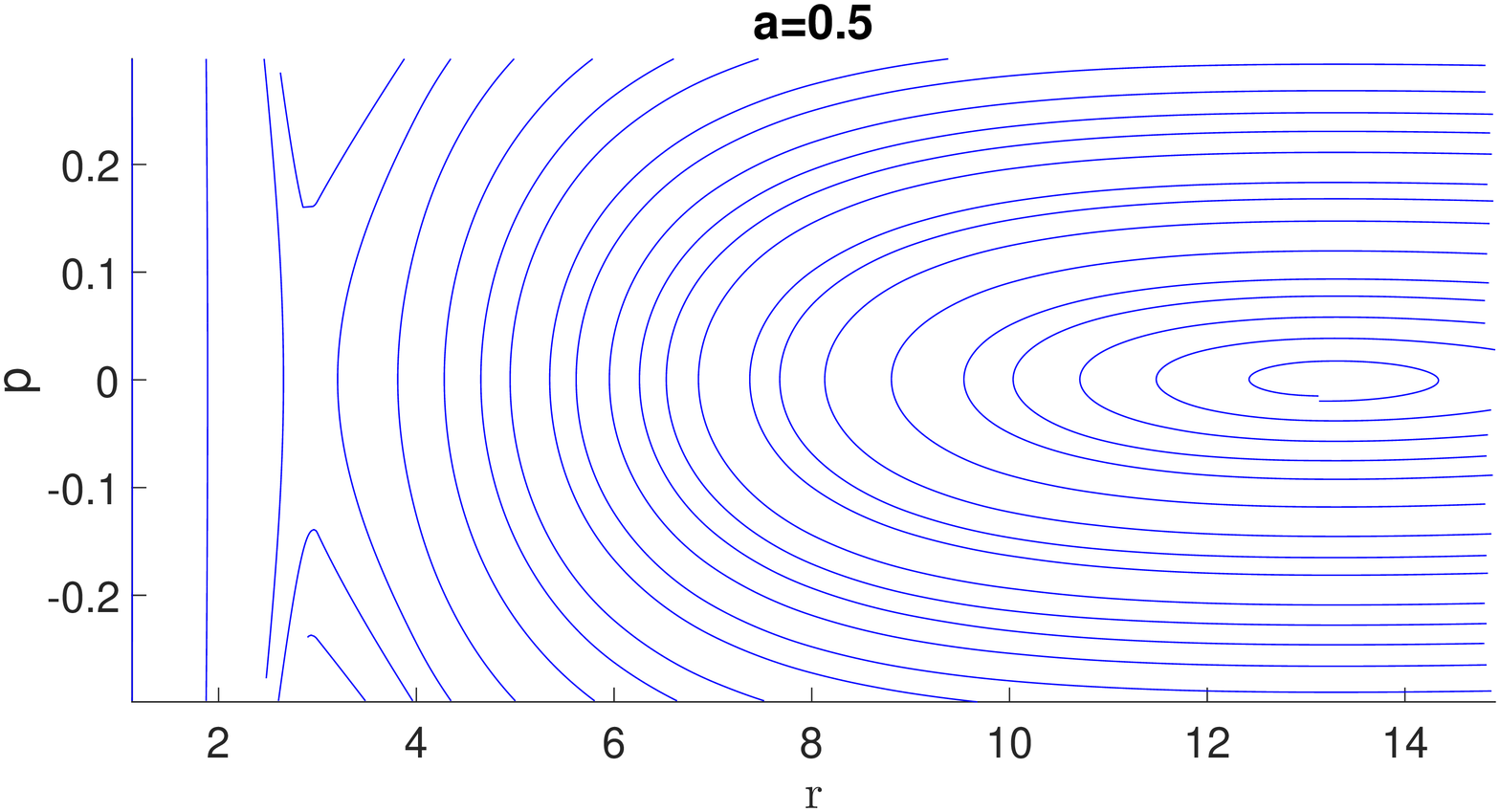}}	\subfigure[]{\includegraphics[width=09cm,height=7.5cm]{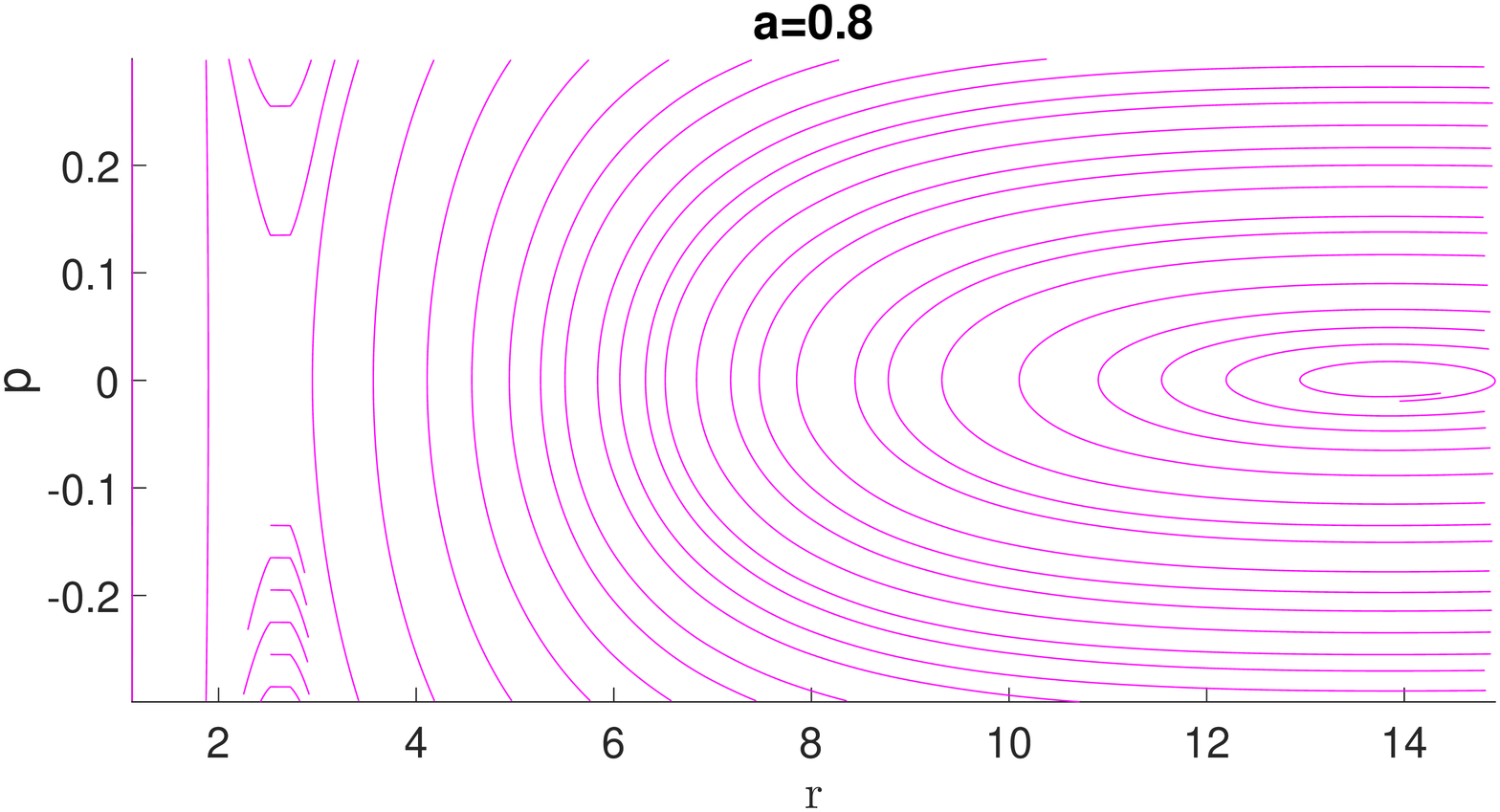}}
	\caption{The $r-p$ phase portrait of Kerr spacetime timelike geodesics for various values of Kerr parameter $'a'$}\label{pptime}
\end{figure}

In this section, we analyze the Lyapunov stability of equilibrium point for timelike geodesics and nature of equilibrium point is shown by corresponding phase-portrait. By substituting  $K=-1$ in the equation \eqref{Kerr effective potential eq} we get the expression of effective potential for timelike geodesics equation as, 
\begin{equation}
V\left(r\right)=\frac{1}{2}-\frac{M}{r}+\frac{l^{2}-a^{2}\left(e^{2}-1\right)}{2r^{2}}-\frac{M\left(l-ae\right)^{2}}{r^{3}} \label{timelike effective potential eq}.
\end{equation}
The derivative of Eq. \eqref{timelike effective potential eq} w.r.t. $'r'$ is obtained as
\begin{equation}
V'\left(r\right)=\frac{M}{r^{2}}-\frac{l^{2}-a^{2}\left(e^{2}-1\right)}{r^{3}}+\frac{3M\left(l-ae\right)^{2}}{r^{4}},
\end{equation}
and the second derivative of Eq.\eqref{timelike effective potential eq} is then
\begin{equation}
V''\left(r\right)=-\frac{2M}{r^{3}}+\frac{3\left(l^{2}-a^{2}\left(e^{2}-1\right)\right)}{r^{4}}-\frac{12M\left(l-ae\right)^{2}}{r^{5}}.
\end{equation}
Therefore, equilibrium points can be obtained by solving the equation $V'(r)=0$, thus we get
\begin{equation}
r_{*}^{\pm}=\frac{l^{2}-a^{2}\left(e^{2}-1\right)\pm\sqrt{\left[l^{2}-a^{2}\left(e^{2}-1\right)\right]^{2}-12M^{2}\left(l-ae\right)^{2}}}{2M}.
\end{equation}
Clearly there are two equilibrium points $\left(r_*^- ,0\right)$ and $\left(r_*^+ ,0\right)$ in the case of timelike geodesics for which $V''\left(r\right)$ is calculated to analyze the Lyapunov stability and shown in the \tablename{\ref{timetable}} for the different values of 'a'.
\begin{table}[ht]
	\caption{Calculation of $V''\left(r\right)$ at the equilibrium point for the timelike geodesics for different values of Kerr parameter 'a' fixing M=1, e=1 and l=4.}
	\centering
	\begin{tabular}{|c|c|c|c|c|c|}
		\hline
		S.No.& a & $r_*^-$& $r_*^+$& $V''(r_*^-)$& $V''(r_*^+)$ \\ 
		[0.7ex]
		\hline
		1  & 0$\left(SBH\right)$   &  4      &  12     & -0.03125 & 0.0003858024  \\
		[0.7ex]
		\hline
		2  & 0.2 & 3.45247 & 12.5475 & -0.0640159 & 0.000366925  \\
		[0.7ex]
		\hline
		3  & 0.5 & 2.77985 & 13.2202 & -0.174834 & 0.00034179   \\
		[0.7ex]
		\hline
		4  & 0.8 & 2.23112 & 13.7689 & -0.465616 & 0.000321014   \\
		[0.7ex]
		\hline
	\end{tabular}  \label{timetable}
\end{table}
\\As, $\lambda=\pm\sqrt{-V''\left(r_{*}\right)}$  therefore from the \tablename{\ref{timetable}}, we can see that at the equilibrium point $\left(r_*^-,0\right)$ the eigenvalues are real, distinct and opposite in sign. So $\left(r_*^-,0\right)$ is a saddle point which is Lyapunov unstable for all the values of specific angular momentum 'a', while at the equilibrium point $\left(r_*^+,0\right)$ the eigenvalues are purely imaginary so $\left(r_*^+,0\right)$ is  center which is Lyapunov stable for all the values of 'a'.\\ To visualize the nature of equilibrium points of timelike geodesics, the phase portraits are depicted for different values of Kerr parameter $'a'$ as shown in \figurename{\ref{pptime}}
\figurename{\ref{pptime}} depicts the phase portrait in $r-p$ plane for stable and unstable circular orbits for timelike geodesics in Kerr spacetime for various values of Kerr parameter $'a'$. For each values of $'a'$, we observe that $V''(r_*^-)<0$ and $V''(r_*^+)>0$ hence equilibrium point $\left(r_*^-,0\right)$ is an unstable saddle point while $\left(r_*^+,0\right)$ is a stable center.\\
The range of stable circular orbit for the different values of $'a'$ in the case of timelike geodesics is given in \tablename{\ref{rangetable}}.
\begin{table*}[ht]
	\caption{Range of stable circular orbits for timelike geodesics for various values of $'a'$} 
	\centering
	\begin{tabular}{|c|c|c|}
		\hline
		S.No. &	a &  Range of stable circular orbits\\ 
		[0.5ex]
		\hline
		1 &	0(SBH)   & $r_*^+>6$ \\
		
		2 &	0.2  & $r_*^+>5.28447$ \\
		
		3 &	0.5 & $r_*^+>4.08432$ \\
		
		5 &	0.8 &  $r_*^+>2.44271$\\
		\hline
	\end{tabular}  
	\label{rangetable}
\end{table*} 
\subsection{For null geodesics}
\begin{figure}[H] 
	\subfigure[]{\includegraphics[width=09cm,height=08cm]{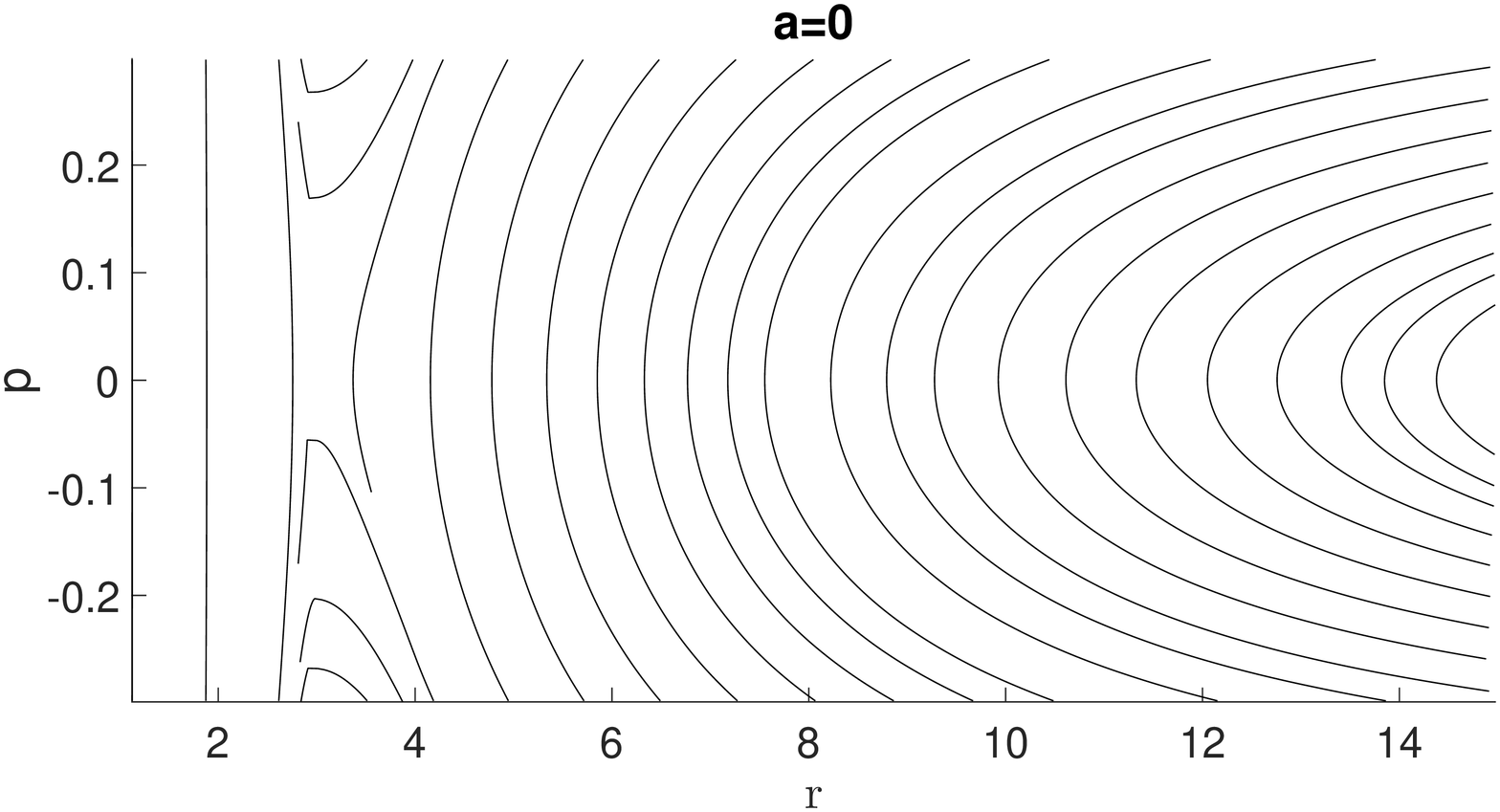}}
	\subfigure[]{\includegraphics[width=09cm,height=08cm]{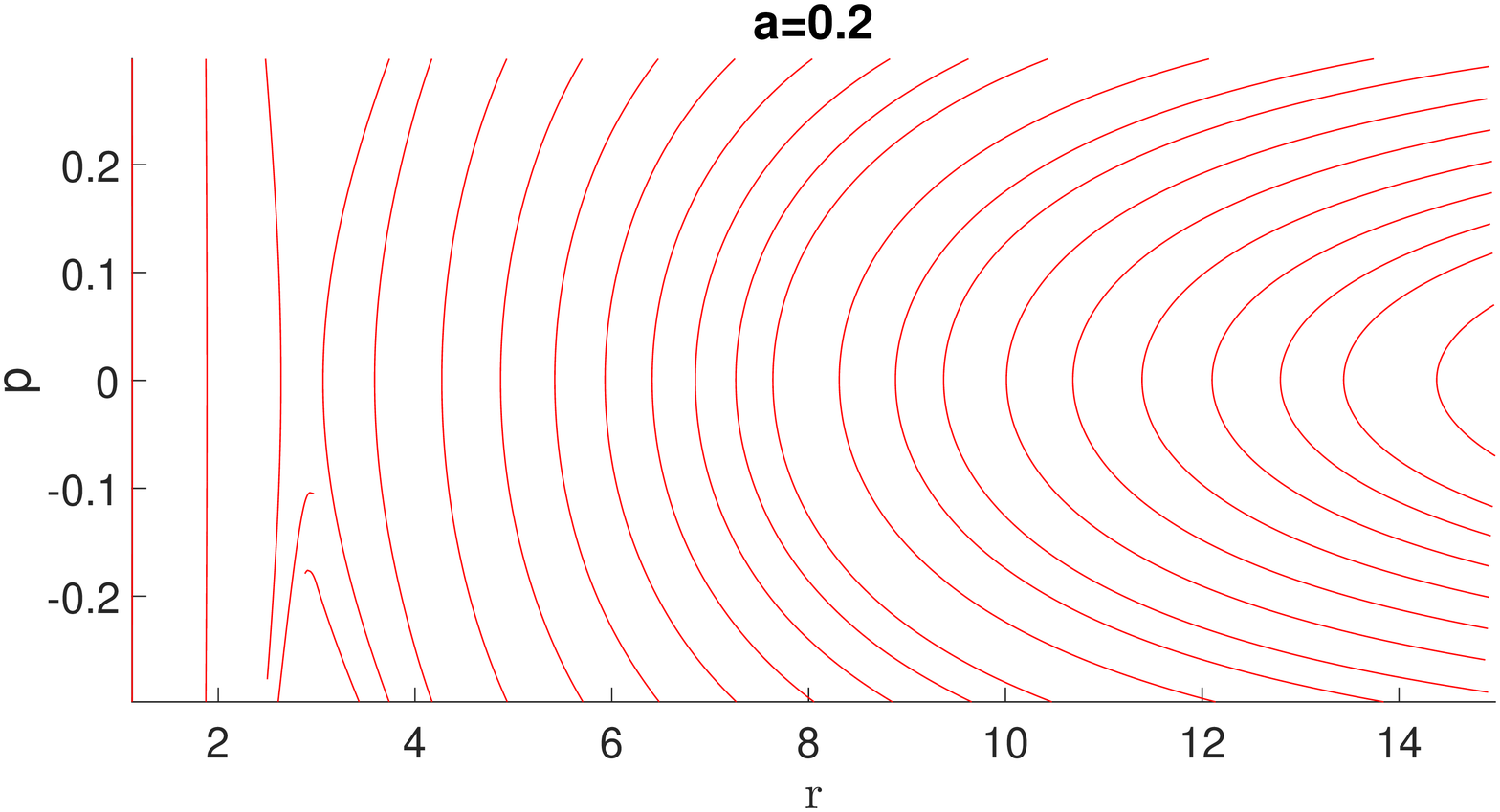}}
	\subfigure[]{\includegraphics[width=09cm,height=08cm]{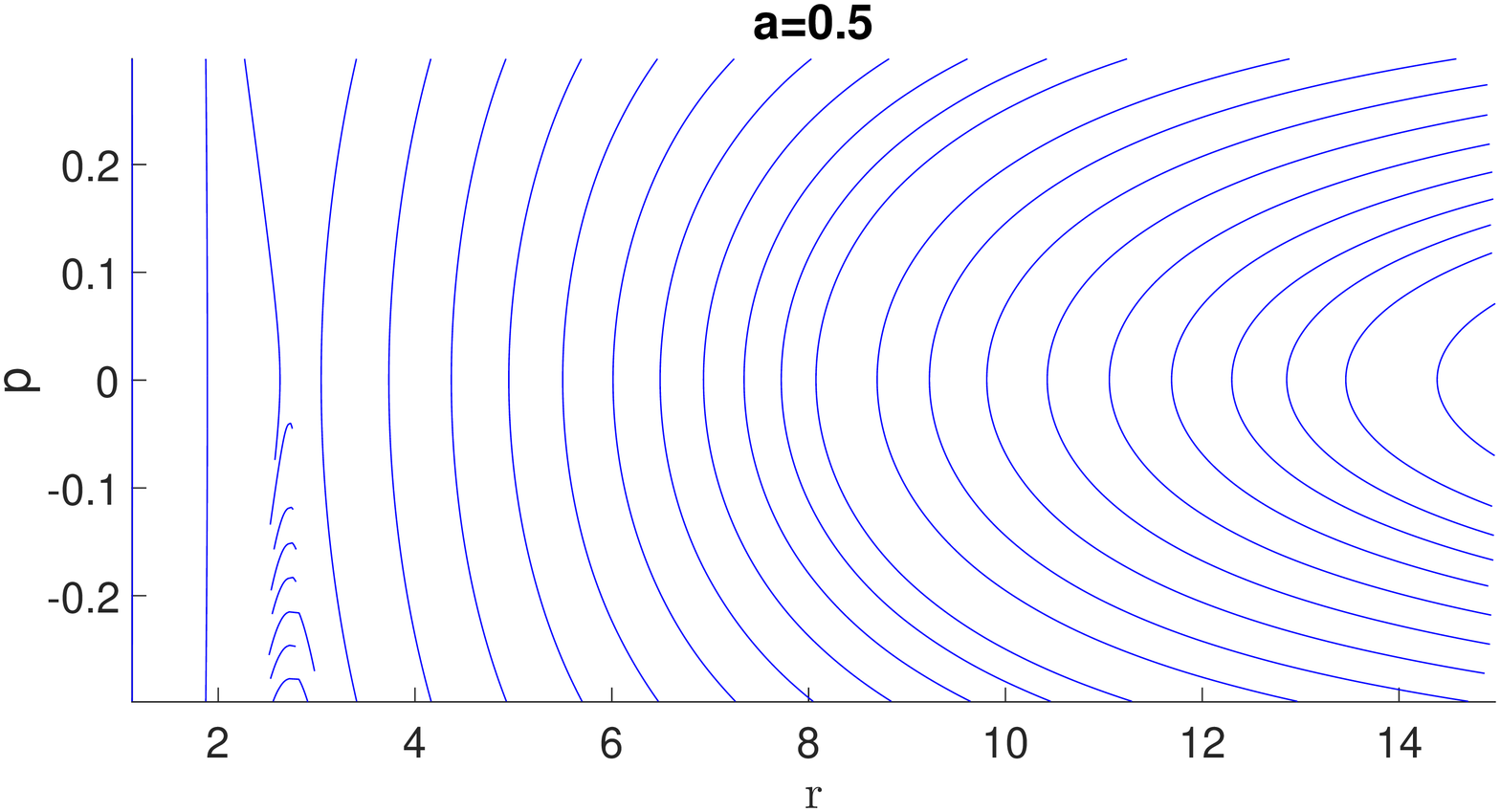}}	\subfigure[]{\includegraphics[width=09cm,height=08  cm]{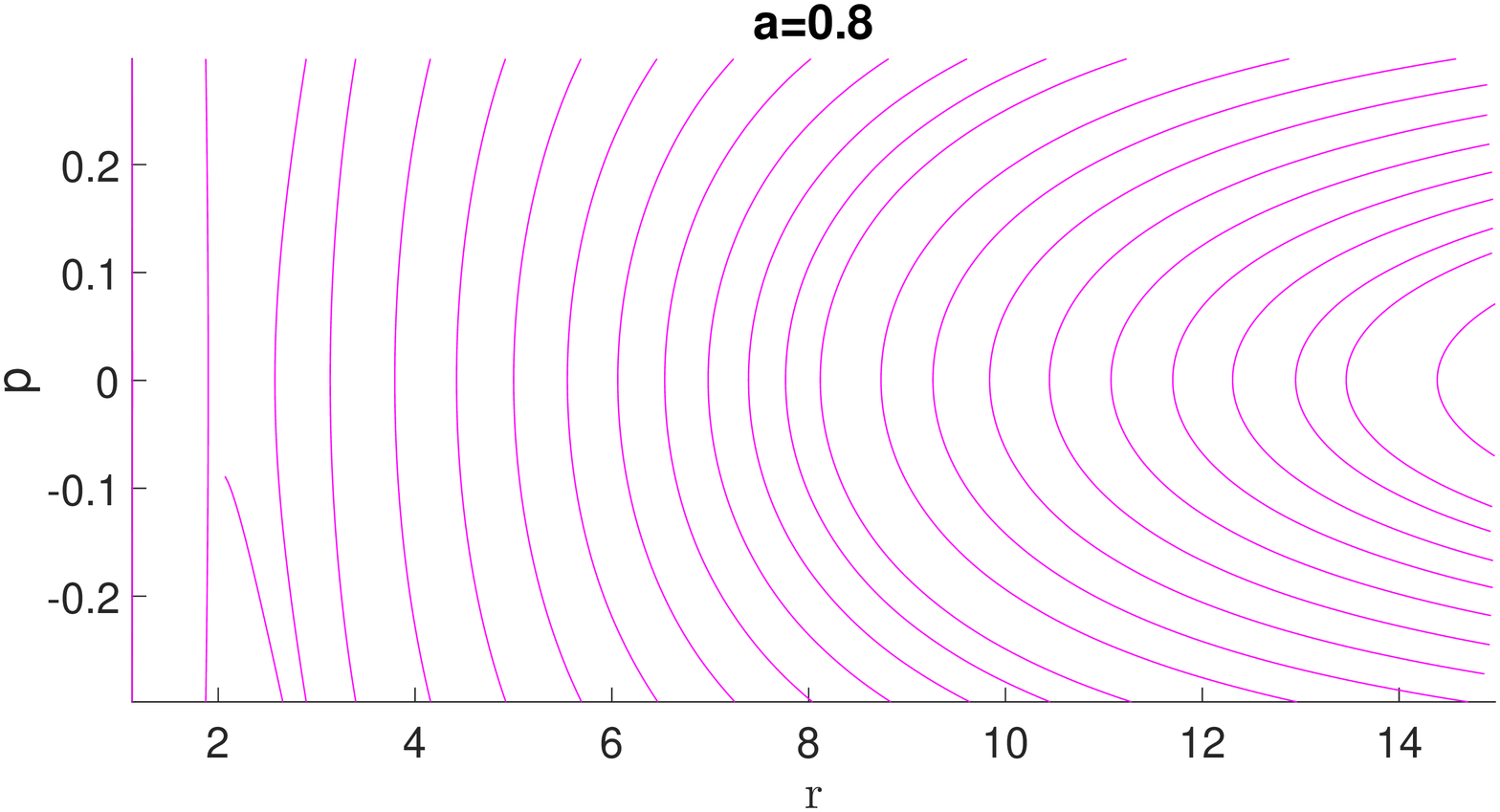}}
	\caption{The $r-p$ phase portrait of  Kerr spacetime null geodesics for various values of Kerr parameter $`a'$.}\label{ppnull}
\end{figure}

By setting $K=0$ in Eq. \eqref{Kerr effective potential eq} the expression of effective potential for null geodesics is obtained as,
\begin{equation}
V\left(r\right)=\frac{l^{2}-a^{2}e^{2}}{2r^{2}}-\frac{M\left(l-ae\right)^{2}}{r^{3}},\label{nulleffpot}
\end{equation}
The derivative of Eq.\eqref{nulleffpot} w.r.t.$`r'$ is given as
\begin{equation}
V'\left(r\right)=-\frac{l^{2}-a^{2}e^{2}}{r^{3}}+\frac{3M\left(l-ae\right)^{2}}{r^{4}},
\end{equation}
and the second derivative is given as
\begin{equation}
V''\left(r\right)=\frac{3\left(l^{2}-a^{2}e^{2}\right)}{r^{4}}-\frac{12\left(l-ae\right)^{2}}{r^{5}}.
\end{equation}
By solving the equation $V'(r)=0$ we obtain
\begin{equation}
r_{*}=\frac{3M\left(l-ae\right)}{l+ae}.
\end{equation}
Here we can see that in the case of null geodesics, we have only one equilibrium point $\left(r_*,0\right)$ for each values of Kerr parameter $`a'$, the value of $V''\left(r\right)$ is calculated and shown in the \tablename{\ref{nulltable}} given below,
\begin{table*}[ht]
	\caption{Calculation of $V''\left(r\right)$ at the equilibrium point for the null geodesics for different values of Kerr parameter $`a'$ when $M=1$, $e=1$ and $l=4$.}
	\centering
	\begin{tabular}{|c|c|c|c|}
		\hline
		S.No, &	a &  $r_*$ & $V''\left(r_*\right)$  \\ 
		[0.7ex]
		\hline
		1  &	0(SBH)   & 3 & -0.197531  \\
		[0.7ex]
		\hline
		2  &	0.2  & 2.71429 & -0.294039 \\
		[0.7ex]
		\hline
		3  &	0.5 & 2.33333 & -0.506045  \\
		[0.7ex]
		\hline
		5  &	0.8 & 2 & -0.96  \\
		[0.7ex]
		\hline
	\end{tabular}  
	\label{nulltable}
\end{table*}
\\From the \tablename{\ref{nulltable}}, since $V''\left(r_*\right)<0$ i.e. $V$ is maximum at $r_*$ therefore from the Eq. \eqref{ch.eq} we see that the eigenvalues of the Jacobian matrix at $\left(r_*,0\right)$ are real, distinct and opposite in sign so the equilibrium point  $\left(r_*,0\right)$ is saddle point which is Lyapunov unstable for all values of $`a'$. Therefore in the case of null geodesics there are no stable circular orbits for any value of Kerr parameter $`a'$.\\ To visualize the complete nature of equilibrium point of null geodesics, phase portrait are depicted in \figurename{\ref{ppnull}} for different values of Kerr parameter $`a'$.
\figurename{\ref{ppnull}} depicts that there is only one extreme point in case of null geodesics for each values of Kerr parameter $`a'$ which is unstable saddle point. No stable circular orbits exist in this case.
\section{Jacobi stability}
The second order differential equation corresponding to the system of Eq. \eqref{one-dim system} is given as 
\begin{equation}
\ddot{r}+V'\left(r\right)=0.
\end{equation}
By using the expression of effective potential of Kerr spacetime given in Eq. \eqref{Kerr effective potential eq},  we get
\begin{equation}
\ddot{r}-\frac{KM}{r^{2}}-\frac{l^{2}-a^{2}\left(e^{2}+K\right)}{r^{3}}+\frac{3M\left(l-ae\right)^{2}}{r^{4}}=0.
\end{equation}
Comparing the above equation with general second order differential equation \eqref{general kcc eq} used in KCC theory we have
\begin{equation}
G^{1}\left(r,p\right)=\frac{1}{2}\left(-\frac{KM}{r^{2}}-\frac{l^{2}-a^{2}\left(e^{2}+K\right)}{r^{3}}+\frac{3M\left(l-ae\right)^{2}}{r^{4}}\right).\label{jacobieq.}
\end{equation} 
The derivative of Eq. \eqref{jacobieq.} with respect to $r$ is obtained as
\begin{equation}
\frac{\partial G^{1}}{\partial  r}=  \frac{1}{2}\Bigg[\frac{2KM}{r^{3}}-\frac{3\left[l^{2}-a^{2}\left(e^{2}+K\right)\right]}{r^{4}}+\frac{12M\left(l-ae\right)^{2}}{r^{5}}\Bigg]\label{diffjacobieq}
\end{equation}
The nonlinear connection associated to this system  is obtained as
\begin{equation}
N^{1}_{1}= \frac{\partial G^{1}}{\partial p}= 0,
\end{equation}
and the Berwald connection is obtained as
\begin{equation}
G^{1}_{11}=\frac{\partial N^{1}_{1}}{\partial p}= 0.
\end{equation}
Finally, the second KCC invariant is given by the equation
\begin{equation}
P^{1}_{1}\left(r,p\right)=-2\frac{\partial G^{1}}{\partial    r}-2G^{1}G^{1}_{11}+p\frac{\partial N^{1}_{1}}{\partial r}+N^{1}_{1}N^{1}_{1}.\label{invarianteq.}
\end{equation}
By substituting the values of $\frac{\partial G^{1}}{\partial  r}$ , $N_{1}^{1}$ and $G_{11}^{1}$ in the Eq.\eqref{invarianteq.} the second KCC invariant is obtained as
\begin{equation}
P^{1}_{1}\left(r,p\right)=-\frac{2KM}{r^{3}}-\frac{3\left[l^{2}-a^{2}\left(e^{2}+K\right)\right]}{r^{4}}+\frac{12M\left(l-ae\right)^{2}}{r^{5}}.
\end{equation}
At the equilibrium point $\left(r_{*},0\right)$ the second KCC invariant is reduced as
\begin{equation}
P^{1}_{1}\left(r_{*},0\right)=-\frac{2KM}{r_{*}^{3}}-\frac{3\left[l^{2}-a^{2}\left(e^{2}+K\right)\right]}{r_{*}^{4}}+  \frac{12M\left(l-ae\right)^{2}}{r_{*}^{5}}.\label{invariant eq.}
\end{equation} 
For the timelike geodesics, inserting $K=-1$ in Eq.\eqref{invariant eq.} we get
\begin{equation}
P^{1}_{1}\left(r_{*},0\right)=\frac{2M}{r_{*}^{3}}-\frac{3\left[l^{2}-a^{2}\left(e^{2}-1\right)\right]}{r_{*}^{4}}+  \frac{12M\left(l-ae\right)^{2}}{r_{*}^{5}}.
\end{equation}
which can be rewritten as following form
\begin{equation}
P^{1}_{1}\left(r_{*},0\right)=\frac{\left(r_*-A\right)\left(r_*-B\right)}{r_*^5},
\end{equation}
where,
\begin{equation}
A=\frac{3\left(l^2-a^2\left(e^2-1\right)\right)-\sqrt{9\left(l^2-a^2\left(e^2-1\right)\right)^2-96M^2\left(l-ae\right)^2}}{4M}, 
\end{equation}
and   
\begin{equation}
B=\frac{3\left(l^2-a^2\left(e^2-1\right)\right)+\sqrt{9\left(l^2-a^2\left(e^2-1\right)\right)^2-96M^2\left(l-ae\right)^2}}{4M}.
\end{equation}       \\ 
Thus, the equilibrium point $\left(r_{*},0\right)$ is Jacobi stable if $P^{1}_{1}\left(r_{*},0\right)<0$\\
i.e.
\begin{equation}
\frac{\left(r_*-A\right)\left(r_*-B\right)}{r_*^5}<0
\end{equation} 
From this inequality, we observed that the point  $\left(r_{*},0\right)$ is Jacobi stable if either $r_*<A$~, ~$r_*>B$~ or ~ $r_*>A$~, ~$r_*<B$, otherwise the point is Jacobi unstable. \\
For the null geodesics, inserting K=0 in Eq.\eqref{invariant eq.} we get
\begin{equation}
P^{1}_{1}\left(r_{*},0\right)=-\frac{3\left(l^{2}-a^{2}e^{2}\right)}{r_{*}^{4}}+  \frac{12M\left(l-ae\right)^{2}}{r_{*}^{5}}.
\end{equation} 
Thus the point $\left(r_{*},0\right)$ is Jacobi stable if $P^{1}_{1}\left(r_{*},0\right)<0$\\ i.e.
\begin{equation}
-\frac{3\left(l^{2}-a^{2}e^{2}\right)}{r_{*}^{4}}+  \frac{12M\left(l-ae\right)^{2}}{r_{*}^{5}}<0
\end{equation}
From this inequality, we obtained that
\begin{equation}
r_*>\frac{4M\left(l-ae\right)}{l+ae}.
\end{equation}
Thus, in the case of null geodesics the equilibrium point  $\left(r_{*},0\right)$ is Jacobi stable if $r_*>\frac{4M\left(l-ae\right)}{l+ae}$, otherwise the point is Jacobi unstable.

\section{Conclusions} 
In this paper, we have studied the stability of timelike as well as null circular geodesics in background of for Kerr BH spacetime on the equatorial plane by using Lyapunov stability and the Jacobi stability analysis. We have analyzed the effect of Kerr parameter (specific angular momentum) $`a'$ of the BH in the region of stable circular orbit by using effective potential and phase portrait analysis. The linear stability analysis is performed by the linearization of the dynamical system via the Jacobian matrix of a non-linear system at the equilibrium point. The KCC theory is used to examine Jacobi stability which shows that the trajectories of the dynamical system bunch together or disperse when approaching the equilibrium point. In the present paper, we see that in the case of timelike geodesics there are two equilibrium points $\left(r_*^-,0\right)$ and $\left(r_*^+,0\right)$ out of which the equilibrium point $\left(r_*^-,0\right)$ is a Lyapunov unstable saddle point  and the another equilibrium point $\left(r_*^+,0\right)$ is stable center.  We also calculate the range of stable circular geodesics for the various values of specific angular momentum $`a'$ for the particle of unit energy per unit mass and observed that as the value of Kerr parameter increases from 0 to 1 the range of stable circular orbits expand as shown in \tablename{\ref{rangetable}}. But in the case of null geodesics there are no stable circular orbits as there is only one equilibrium point $\left(r_*,0\right)$ which is a Lyapunov unstable saddle point.  Further, we investigate the Jacobi stability for timelike as well as null geodesics and obtained the condition for Jacobi stable equilibrium point. In the case of timelike geodesics, the equilibrium point $\left(r_*,0\right)$ is Jacobi stable if either $r_*<A$~ , ~$r_*>B$~ or ~ $r_*>A$~ , ~$r_*<B$, otherwise the point is Jacobi unstable and in the case of null geodesics, the  equilibrium point $\left(r_*,0\right)$ is Jacobi stable if $r_*>\frac{4M\left(l-ae\right)}{l+ae}$, otherwise the point is Jacobi unstable.

\section*{\normalsize Acknowledgments}
{\normalsize  P.S. would like to thank University Grants Commission $\left(UGC\right)$, New Delhi, India for providing the financial support as a Junior Research Fellow through UGC-Ref.No. 1060/CSIR-UGC NET-JUNE2018. H.N. thankfully acknowledges the financial support provided by Science and Engineering Research Board (SERB), India through grant no. EMR/2017/000339. The authors also acknowledge the facilities available at ICARD, Gurukula Kangri (Deemed to be University) Haridwar those were used during the course of this work.  }  

\bibliographystyle{unsrt} 
\bibliography{PradeeRef} 
\end{document}